\documentstyle[12pt]{article}
\input amssym.def
\input amssym.tex

\def\goth{\frak}
\def\double{\Bbb}
\def\ccal{\cal}

\def\cc{{\double C}}

\def\rr{{\double R}}
\def\zz{{\double Z}}
\def\qqq{{\double Q}}

\def\aa{{\cal A}}

\def\dd{{\cal D}}
\def\gg{{\goth g}}
\def\hh{{\cal H}}
\def\hhh{{{\double H}}}
\def\ff{{\cal F}}
\def\mm{{{\ccal M}}}

\def\aa{{\cal A}}
\def\dd{{\cal D}}
\def\hh{{\cal H}}
\def\ff{{\cal F}}

\def\sss{{\cal S}}
\def\t{\,{\rm tr}\,}

\def\ddd{{\,\hbox{$\partial\!\!\!/$}}}
\def\dee{\,\hbox{\rm D}}
\def\de{\,\hbox{\rm d}}

\def\lb{\left[}
\def\rb{\right]}

\def\ul{\underline}
\def\ot{\otimes}
\def\op{\oplus}

\def\bb{\begin{eqnarray}}
\def\ee{\end{eqnarray}}
\def\eee{\nonumber\end{eqnarray}}
\def\pp{\pmatrix}
\def\qq{\quad}

\begin{document}

\hsize 17truecm
\vsize 24truecm
\font\twelve=cmbx10 at 13pt
\font\eightrm=cmr8
\baselineskip 18pt

\begin{titlepage}

\centerline{\twelve CENTRE DE PHYSIQUE THEORIQUE}
\centerline{\twelve CNRS - Luminy, Case 907}
\centerline{\twelve 13288 Marseille Cedex}
\vskip 4truecm

\centerline{\twelve  Fuzzy Mass Relations for the
Higgs}

\bigskip

\begin{center}
{\bf Bruno IOCHUM}
\footnote{ and Universit\'e de Provence,
\qq\qq iochum@cpt.univ-mrs.fr
\qq schucker@cpt.univ-mrs.fr} \\
\bf Daniel KASTLER
\footnote{ and Universit\'e d'Aix-Marseille II} \\
\bf Thomas SCH\"UCKER $^{1}$
\end{center}

\vskip 2truecm
\leftskip=1cm
\rightskip=1cm
\centerline{\bf Abstract}

\medskip

The non-commutative approach of the standard model
produces a relation between the top and the Higgs
masses. We show that, for a given top mass,
the Higgs mass is constrained to lie in an interval. The
length of this interval is of the order of
$m_\tau^2/m_t$.

\vskip 1truecm
PACS-92: 11.15 Gauge field theories\\
\indent
MSC-91: 81E13 Yang-Mills and other gauge theories

\vskip 2truecm

\noindent april 1995
\vskip 1truecm
\noindent CPT-95/P.33197\\

\vskip1truecm

 \end{titlepage}

We believe that non-commutative geometry \cite{c} is
about to revolutionize physics \cite{c}, \cite{cl} to the
same extent as Minkowskian and Riemannian
geometry did. A first clear prediction of the new
theory is the value of the Higgs mass \cite{ks2},
$m_H=280\pm 33\ GeV,$ if
 the top
mass is $m_t=176\pm18\ GeV$. The aim of this review
is to appreciate the concepts behind this numerical
constraint.

Let us view a Yang-Mills-Higgs model as a point in an
infinite discrete space and a real parameter space.
The points are labeled by an arbitrary finite
dimensional, real, compact Lie group $G$, three
arbitrary unitary representations $\rho_L,\ \rho_R,\
\rho_S$ and several multi-linear invariants of order
two, three and four. The group describes the gauge
interactions. The representations describe the
spectrum of left- and right-handed fermions and of
scalars.  The invariants are parameterized by gauge
couplings, Yukawa couplings and scalar
self-couplings and the parameter space is some
Cartesian power of the real line. This power depends
on the point in the discrete space.  Today's dilemma of
particle physics can be summarized as follows:
Experiment has singled out a mediocre point in the
discrete space, the standard model. Its real parameter
space is 18 dimensional and without any structure.
Namely, the standard model contains 3 gauge
couplings,  masses for the $W$, 3 leptons, 6 quarks, 1
scalar and 4 Cabbibo-Kobayashi-Maskawa mixing
parameters, that all remain arbitrary.

In the non-commutative Connes-Lott approach to
Yang-Mills-Higgs models, the entire Higgs sector
comes free of charge. Thereby both the discrete and
the real parameter space is reduced tremendously
\cite{is2}. While the two Yang-Mills-Higgs spaces are
hypercubes, both Connes-Lott spaces have a rich
structure. In particular, the Higgs mass is forces into
an interval. The length is of the order of
$m_\tau^2/m_t$.

\section{ Yang-Mills-Higgs models}

Let us first set up our notations of a YMH model. It is
defined by the following input:
\begin{itemize}
\item
 a finite dimensional, real, compact Lie group $G$,
\item
 an invariant scalar product
 on the Lie algebra  $\gg$ of $G$, this
choice being parameterized by a few positive numbers
$g_i$, the gauge coupling,
\item
a (unitary) representation $\rho_L$ on a Hilbert space
$\hh_L$ accommodating the left handed fermions
$\psi_L$,
  \item
a representation $\rho_R$ on $\hh_R$ for the right
handed fermions $\psi_R$,
\item
 a representation $\rho_S$ on $\hh_S$ for the scalars
$\varphi$,  \item
 an invariant,
positive polynomial $V(\varphi),\  \varphi \in \hh_S$
 of order 4, the Higgs potential,
\item
 one complex number or Yukawa
coupling $g_Y$ for every trilinear invariant, i.e. for
every one dimensional invariant subspace, ``singlet'', in
the decomposition of the representation  associated to
$  \left(\hh_L^{\ast}\otimes
\hh_R\otimes \hh_S\right) \oplus\left(
     \hh_L^\ast\otimes \hh_R\otimes \hh_S^*\right).
$
\end{itemize}
The standard model is defined by the following input:
\bb G =  SU(3) \times SU(2) \times U(1) \eee
with three coupling constants $ g_3, g_2, g_1 $
defined conventionally by
\bb (b_1,b_2)&:=&\frac{1}{g_1^2}\bar b_1b_2,
\qq b_1,b_2\in\ u(1),
\cr\cr  (a_1,a_2)&:=&\frac{2}{g_n^2}\t(a^*_2a_2),\qq
a_1,a_2\in su(n). \eee
The representations are
\bb
\hh_L &=& \bigoplus_1^3\lb (1,2,-1)\oplus (3,2,{1
\over 3}) \rb  \label{HL},\cr \cr
 \hh_R& = &\bigoplus_1^3\lb (1,1,-2)\oplus
(3,1,{4\over 3})\oplus (3,1,-{2\over 3}) \rb, \cr\cr
 \hh_S &= &(1,2,-1/2) ,\eee
where $(n_3,n_2,y)$
denotes the tensor product of an $n_3$ dimensional
representation of $SU(3)$, an $n_2$ dimensional
representation of $SU(2)$ and the one dimensional
representation of $U(1)$ with hypercharge $y$:
\bb
\rho(e^{i\theta}) = e^{iy\theta}, &&\qquad
y\in\qqq,\   \theta \in [0,2\pi),\cr
 V(\varphi) = \lambda (\varphi^\ast
\varphi)^2 - {{\mu ^2 }\over 2}\varphi^\ast \varphi,
&&\qquad \varphi \in
\hh_S, \quad \lambda, \mu > 0.\eee
There are 27 Yukawa couplings in the standard
model, that can be traded in for 9 fermion masses
and 4 mixing parameters.

The gauge symmetry is said to be spontaneously broken
 if every minimum $v\in \hh_S$ of the
Higgs potential is gauge variant,
$ \rho_S(g)v\ne v$ for some $ g\in G$.
Any such minimum $v$ is called a vacuum. For
simplicity let us assume that the vacuum is
non-degenerate, i.e. all minima lie on the same orbit
under $G$.
To do perturbation theory, we have to introduce a scalar
variable $h$, that vanishes in the vacuum,
\bb h(x):=\varphi(x)-v,\eee
$x$ a point in spacetime $M$.
With this change of variables, the Klein-Gordon
Lagrangian is $ \left({\rm
D}\varphi\right)^**{\rm D} \varphi$.
 The Hodge star $*\cdot$
should be distinguished from the Hilbert space dual
$\cdot^*$, wedge symbols are suppressed.
 We denote by D
the covariant exterior derivative, here for scalars
$
\dee\varphi:=\de\varphi+\tilde\rho_S(A)\varphi,$
$\varphi$ is now a multiplet of {\it fields}, i.e. a 0-form
on spacetime with values in the scalar representation
space,
$ \varphi\in\Omega^0(M,\hh_S),$
while the vacuum $v$ remains  constant over
spacetime so that it also minimizes the kinetic term
$\de\varphi^**\de\varphi$. The gauge fields are
1-forms with values in the Lie algebra of $G$:
$ A\in\Omega^1(M,\gg),$
$\tilde\rho_S$ denotes the Lie algebra representation
on $\hh_S$. The Klein-Gordon Lagrangian produces
the mass matrix for the gauge bosons $A$.
This mass matrix is given by the (constant) symmetric,
positive semi-definite form on the Lie algebra of $G$,
\bb \left(\tilde\rho_S(A)v\right)^*\tilde\rho_S(A)v.
\eee
It contains the masses of the gauge bosons some of
which remain massless. In the example of the standard
model, these are the photon and the gluons.

In the following we are more concerned with the
fermionic mass matrix $\mm$, a linear map $ \mm:
\hh_R \longrightarrow \hh_L$. We want to produce it
in the same way we produced the mass matrix for the
gauge bosons, via the change of variables
$h(x):=\varphi(x)-v$. For this purpose, we add by hand
to the Dirac Lagrangian gauge invariant trilinears
\bb \sum_{j=1}^ng_{Yj}\left(\psi_L^*,\psi_R,\varphi
\right)_j+\sum_{j=n+1}^mg_{Yj}\left(\psi_L^*,\psi_R,
\varphi^*\right)_j\ +\ {\rm complex\ conjugate}
\label{tri},\ee
$n$ is the number of singlets in
$\left(\hh_L^{\ast}\otimes\hh_R\otimes \hh_S\right)
$, $m+n$ the number of singlets in
$\left(
     \hh_L^\ast\otimes \hh_R\otimes \hh_S^*\right)$.
For $h=0$ again, we
obtain the fermionic mass matrix $\mm$ as a function
of the Yukawa couplings $g_{Yj}$ and the vacuum $v$
\bb \psi_L^*\mm\psi_R:         =
\sum_{j=1}^ng_{Yj}\left(\psi_L^*,\psi_R,v
\right)_j+\sum_{j=n+1}^mg_{Yj}\left(\psi_L^*,\psi_R,
v^*\right)_j.\eee
 As
the gauge boson masses, the fermionic mass terms
$\psi_L^*\mm\psi_R$ are not gauge invariant in
general. They are gauge invariant if
$ \rho_L(g^{-1})\mm\rho_R(g)=\mm\
{\rm for\ all}\ g\in G$.
In the standard model with its 27 Yukawa couplings,
the mass matrix $\mm$ can be any matrix yielding
mass terms invariant under $SU(3)\times U(1)$.

\section{ Connes-Lott models}

This section summarizes the non-commutative
approach to Yang-Mills-Higgs models,
 \cite{c}, \cite{cl},  \cite{k}, \cite{vg}. Although we
shall follow this approach due to Connes and Lott, let us
mention that there are alternative approaches
similar in spirit,  \cite{orsay}, \cite{mm},  \cite{zur}.

\subsection{Internal space}

A Connes-Lott model is defined by the
following choices:
\begin{itemize}
\item
 a finite dimensional, associative,
algebra $\aa$ over the field $\rr$ or $\cc$ with unit 1
and involution  \nolinebreak $^*$,
\item
 two *- representations of $\aa$, $\rho_L$ and
$\rho_R$, on Hilbert spaces $\hh_L$ and $\hh_R$
over the field, such
that $\rho:=\rho_L\op\rho_R$ is faithful,
\item
 a mass matrix $\mm$ i.e. a linear map
$ \mm: \hh_R \longrightarrow\hh_L, $
\item
a certain number of
coupling constants depending on the degree of
reducibility of $ \rho_L\oplus \rho_R. $
\end{itemize}
The data $(\hh_L,\hh_R,\mm)$ plays
a fundamental role in non-commutative Riemannian
geometry where it is called K-cycle.

With this input and the rules of non-commutatative
geometry, Connes and Lott construct a YMH model.
Their starting point is
an auxiliary differential algebra $\Omega \aa,$
the so called universal differential envelope of $\aa$:
\bb   \Omega^0\aa := \aa,\eee
$\Omega^1\aa$ is
generated by symbols $\delta a$, $ a \in \aa$ with
relations
\bb   \delta 1 = 0,\qq     \delta(ab) = (\delta
a)b+a\delta b.\eee
 Therefore $\Omega^1\aa$ consists
of finite sums of terms of the form $a_0\delta a_1,$
\bb   \Omega^1\aa = \left\{ \sum_j a^j_0\delta
a^j_1,\quad a^j_0, a^j_1\in \aa\right\}\eee
and likewise
for higher $p$,
\bb   \Omega^p\aa = \left\{ \sum_j
a^j_0\delta a^j_1...\delta a^j_p ,\quad a^j_q\in
\aa\right\}.\eee
 The differential $\delta$ is defined by
$   \delta(a_0\delta a_1...\delta a_p) :=
   \delta a_0\delta a_1...\delta a_p.$

Two remarks: The universal differential envelope
$\Omega\aa$ of a commutative algebra $\aa$ is
not necessarily graded commutative. The
universal differential envelope of any algebra has no
cohomology. This means that every closed form
$\omega$ of degree  $p\geq 1,\quad
\delta\omega=0,$ is exact, $\omega =
\delta\kappa$ for some $(p-1)$form $\kappa.$

The involution $^*$ is
extended from the algebra $\aa$ to
 $\Omega^1\aa$ by putting
\bb   (\delta a)^* := \delta(a^*) =:\delta a^*.\eee
 With the definition
$ (\omega\kappa)^*=\kappa^*\omega^*$,
the involution is extended to the whole
differential envelope.

The next step is to extend the
representation $\rho:=\rho_L\op\rho_R$ on
$\hh:=\hh_L\op\hh_R$ from the  algebra $\aa$   to its
universal differential envelope $\Omega\aa$. This
extension is the central piece of Connes' algorithm
and deserves a new name:
\bb && \pi :
\Omega\aa  \longrightarrow{\rm End}(\hh)
   \cr
 &&\pi(a_0\delta a_1...\delta a_p) :=
(-i)^p\rho(a_0)[\dd,\rho(a_1)] ...[\dd,\rho(a_p)]
\eee
where $\dd$ is the linear map from
$\hh$ into itself
\bb   {\dd}:= \pmatrix {0 & \mm \cr\mm^*
&0}.\eee
In non-commutative geometry, $\dd$ plays the role of
the Dirac operator and we call it internal
Dirac operator. A
straightforward calculation shows that $\pi$ is in fact
a  representation of $\Omega\aa$ as involution
algebra, and we are tempted to define also a
differential, again denoted by $\delta$,
 on $\pi(\Omega\aa)$ by
\bb
\delta\pi(\omega):=\pi(\delta\omega).
\label{trial}\ee
However, this definition does not make sense if there
are forms
$\omega\in\Omega\aa$ with
$\pi(\omega)=0$ and  $\pi(\delta\omega)
\not= 0$. By dividing out these unpleasant forms,
Connes constructs a new differential algebra
$\Omega_\dd\aa$, the  interesting object:
\bb   \Omega_\dd\aa :=
{{\pi\left(\Omega\aa\right)}\over J}\eee
with
\bb   J := \pi\left(\delta\ker\pi\right) =:
\bigoplus_p J^p\eee
($J$ for junk).
On the quotient now,
the differential (\ref{trial}) is well defined. Degree by
degree we have:  \bb   \Omega_\dd^0\aa =\rho(\aa)\eee
because $J^0=0$ ,
\bb   \Omega_\dd^1\aa = \pi(\Omega^1\aa)\eee
because $\rho$ is faithful,
and in degree $p\geq2$,
\bb   \Omega_\dd^p\aa =
{{\pi(\Omega^p\aa)}\over
{\pi(\delta(\ker\pi)^{p-1})}}.\eee
While $\Omega\aa$ has no cohomology,
$\Omega_\dd\aa$ does in general. In fact, in infinite
dimensions,
if $\ff$ is the algebra of complex functions on
spacetime $M$ and if the K-cycle is
obtained from the Dirac operator then
$\Omega_\ddd\ff$ is de Rham's differential algebra of
differential forms on $M$.

We come back to our finite
dimensional case. Remember that the elements of the
auxiliary differential algebra $\Omega\aa$ that
we introduced for book keeping purposes only, are
abstract entities defined in terms of symbols and
relations. On the other hand the elements of
$\Omega_\dd\aa$, the ``forms'', are operators on the Hilbert
space $\hh$, i.e. concrete matrices of
complex numbers. Therefore there is a
natural scalar product defined by
\bb   <\hat\omega,\hat\kappa> :=
\t (\hat\omega^*\hat\kappa),
\quad  \hat\omega, \hat\kappa \in
\pi(\Omega^p\aa)\label{sp}\ee
for elements of equal degree and by zero for two
elements of different degree.
With this scalar product  $\Omega_\dd\aa$ is a
subspace of $\pi(\Omega\aa)$, by  definition
orthogonal to the junk. As a subspace,
$\Omega_\dd\aa$  inherits a scalar product which deserves
a special name ( , ). It is  given by
\bb   (\omega,\kappa) =
<\hat\omega,P\hat\kappa>, \quad \omega, \kappa
\in \Omega_\dd^p\aa\eee
 where $P$ is the orthogonal
projector in $\pi(\Omega\aa)$  onto the
ortho-complement of $J$ and $\hat\omega$ and
$\hat\kappa$ are any  representatives in the classes
$\omega$ and $\kappa$. Again the scalar product
vanishes  for forms with different degree. For real
algebras, all traces must be understood as real part of
the trace.

In Yang-Mills models coupling constants appear as
parameterization of the most general gauge invariant
scalar product. In the same spirit, we want the most
general scalar product on $\pi(\Omega\aa)$
compatible with the underlying algebraic structure. It
is given by
\bb   <\hat\omega,\hat\kappa>_z :=
\t (\hat\omega^*\hat\kappa\,z),
\quad  \hat\omega, \hat\kappa \in
\pi(\Omega^p\aa),\label{gsp}\ee
where $z$ is a positive operator on $\hh$, that
commutes with $\rho(\aa)$, with the Dirac
operator $\dd$ and with the chirality operator $\chi$,
\bb \chi\psi_L=-\psi_L,\qq\chi\psi_R=+\psi_R.\eee
 A natural subclass of these scalar products
is constructed with operators $z$ in the image under
$\rho$ of the center of $\aa$.

Since $\pi$ is a homomorphism of involution algebras
the product  in $\Omega_\dd\aa$ is given by matrix
multiplication followed by the  projection $P$.
The involution  is
given by transposition and complex conjugation, i.e.
the dual with respect to the scalar product of the Hilbert
space $\hh$. Note that this scalar product admits no
generalization. W. Kalau et al. \cite{kppw} discuss the
computation of the junk and of the differential for
matrix algebras.

At this stage there is a first contact with gauge
theories.  Consider the vector space of anti-Hermitian
1-forms
$   \left\{ H\in \Omega_\dd^1\aa,\ H^*=-H
\right\}.$
A general element $H$ is of the form
\bb   H =
i\pmatrix {0&h\cr h^*&0}\eee
with $h$ a finite sum of terms
$\rho_L(a_0)[\rho_L(a_1)\mm-\mm\rho_R(a_1)]:\
\hh_R \rightarrow \hh_L,\  a_0,a_1\in\aa.$
 These elements are
called Higgses or gauge potentials.  In fact the space of
gauge potentials carries an affine  representation of
the group of unitaries
\bb    G: = \{g\in \aa,\ gg^\ast
=g^*g=1\}\eee
defined by
\bb H^g &:=&\
\rho(g)H\rho(g^{-1})+\rho(g)\delta (\rho(g^{-1})) \cr
       &=&\ \rho(g)H\rho(g^{-1})+(-i)\rho(g)[\dd,\rho
(g^{-1})] \cr
      &=&\ \rho(g)[H-i\dd]\rho(g^{-1})+i\dd\cr\cr
       &=& i\pmatrix {0&h^g\cr (h^g)^*&0}\eee
with
$    h^g-\mm :=
\rho_L(g)[h-\mm]\rho_R(g^{-1})$.
 $H^g$ is
the ``gauge transformed of $H$''.  As usual every gauge
potential $H$ defines a covariant derivative  $\delta
+H$, covariant under the left action of $G$ on
$\Omega_\dd\aa$:
\bb   ^g\omega := \rho(g)\omega, \quad
\omega\in\Omega_\dd\aa\eee
 which means
\bb   (\delta+H^g)\
^g\omega = \ ^g\lb(\delta+H)\omega\rb.\eee
 Also we define the
curvature $C$ of $H$ by
\bb   C := \delta H+H^2\ \in\Omega_\dd^2\aa.\eee
Note that here and later, $H^2$ is considered as element
of $\Omega_\dd^2\aa$ which means it is the projection $P$
applied to $H^2\in \pi(\Omega^2\aa)$.
The curvature $C$ is a Hermitian 2-form with {\it
homogeneous} gauge transformations
 \bb   C^g :=
\delta(H^g)+(H^g)^2 = \rho(g) C \rho(g^{-1}).\eee
Finally, we define the preliminary Higgs potential
$V_0(H)$, a functional on the  space of gauge
potentials, by
\bb   V_0(H)
:= (C,C) = \t[(\delta H+H^2)P(\delta H+H^2)].\eee
It is a
polynomial of degree 4 in $H$ with real, non-negative
values.  Furthermore it is gauge invariant,
$V_0(H^g)= V_0(H)$,
 because of the homogeneous transformation
property of the  curvature $C$ and because the
orthogonal projector $P$ commutes  with all gauge
transformations,
$\rho(g)P =P\rho(g)$.
The most remarkable property of the preliminary
Higgs potential is that, in most cases, its vacuum
spontaneously breaks the group $G$. To simplify the
discussion, let us assume that the Dirac operator is a
1-form,
\bb \dd\in\Omega_\dd^1\aa.\label{hy}\ee
Models not satisfying this hypothesis typically have
degenerate vacua \cite{is2}.
Then, we can introduce the change of variables
\bb\Phi:=  H-i\dd=:i\pmatrix {0&\varphi \cr
\varphi^*&0}\in\Omega_\dd^1\aa\label{fi}\ee
with $\varphi=h-\mm$. Assuming of course a
gauge invariant internal Dirac operator, $\dd^g=\dd$,
$\Phi$ is homogeneously transformed into
\bb\Phi^g&=&H^g-\dd
=\rho(g)[H-i\dd]\rho(g^{-1})+i\dd-i\dd
=\rho(g)\Phi\rho(g^{-1}),\label{hom}\ee
and \bb\varphi^g =
\rho_L(g)\varphi\rho_R(g^{-1}).\eee
Now $h=0$, or
equivalently $\varphi=-\mm$, is certainly a
minimum of the preliminary Higgs potential and this
minimum spontaneously breaks $G$ if it is gauge
variant and non-degenerate.

Consider two extreme classes of examples,
vector-like and left-right models.

A {\it vector-like model} is defined by an arbitrary
internal algebra $\aa$ with identical left and right
representations, $\rho_L=\rho_R$, and with a mass
matrix proportional to the identity in each irreducible
component. Since $\dd$ and $\rho$
commute, the internal differential algebra is trivial,
$\Omega_\dd^p\aa=0$ for $p\geq 1$ and the Dirac
operator is not a
1-form,  $\dd\not\in\Omega_\dd^1\aa$.
However, as we shall see, every vector-like model
produces a Yang-Mills model with unbroken parity and
unbroken gauge symmetry
 as electromagnetism or
chromodynamics.

We define a {\it left-right model} by an internal
algebra consisting of a sum of  a ``left-handed'' and a
``right-handed'' algebra, $\aa=\aa_L\op\aa_R$ with the
left-handed algebra acting only on left-handed
fermions and similarly for right-handed
\bb
\rho_L(a_L,a_R)=\rho_L(a_L,0),\quad
\rho_R(a_L,a_R)=\rho_R(0,a_R),\qq
a_L\in\aa_L, \ a_R\in\aa_R.\eee
Now, any non-vanishing fermion mass
matrix breaks the gauge invariance. At the same time,
the internal Dirac operator is always a 1-form, $\dd
\in\Omega_\dd^1\aa$.

\subsection{Adding spacetime}

 In the next step, the vectors $\psi_L,\ \psi_R$, and $H$
are promoted to genuine fields, i.e. rendered spacetime
dependent. As already in classical quantum mechanics,
this is achieved by tensorizing with functions. Let us
denote by $\ff$ the algebra of
(smooth, real or complex valued) functions over
spacetime $M$. Consider the algebra $\aa_t:=\ff\ot\aa$.
The group of unitaries of the tensor algebra $\aa_t$ is
the gauged version of the group of unitaries of the
internal algebra $\aa$, i.e. the group of functions
from spacetime into the group $G$. Consider the
representation $\rho_t:=\ul\cdot\ot\rho$ of the
tensor algebra on the tensor product
$\hh_t:=\sss\ot\hh,$
where $\sss$ is the Hilbert space of square integrable
spinors on which functions act by multiplication:
$ (\ul f\psi)(x):=f(x)\psi(x)$, $ f\in\ff,\
\psi\in\sss$.
The definition of the tensor product of Dirac operators,
\bb\dd_t:=\ddd\ot 1+\gamma_5\ot\dd\eee
comes from non-commutative geometry. We now
repeat the above construction for the infinite
dimensional algebra $\aa_t$ and its K-cycle. As
already stated, for $\aa=\cc,\ \hh=\cc,\ \mm=0$ the
differential algebra $\Omega_{\dd_t}\aa_t$ is
isomorphic to the de Rham algebra of differential
forms $\Omega (M,\cc)$. For general $\aa$, using the
notations of  \cite{sz}, an anti-Hermitian 1-form
$ H_t\in\Omega_{\dd_t}^1\aa_t,$
\bb H_t=A+H,\eee
 contains two pieces, an
anti-Hermitian Higgs {\it field} $
H\in\Omega^0(M,\Omega_\dd^1\aa)$ and a genuine
gauge field $ A\in\Omega^1(M,\rho(\gg))$
with values in the Lie algebra of the group of
unitaries,
 $ \gg:=\left\{ X\in\aa,\ X^*=-X\right\},$
represented on $\hh$. The curvature of $H_t$
\bb C_t:=\delta_tH_t+H_t^2\ \in\Omega_\dd^2\aa_t\eee
contains three pieces,
\bb C_t=C+F-\dee\Phi\gamma_5,\eee
 the ordinary, now $x$-dependent,
curvature $C=\delta H+H^2$, the field strength
\bb F:=\de A+\frac{1}{2}[A,A]\quad \in
\Omega^2(M,\rho(\gg))\eee
and the covariant derivative of $\Phi$
\bb\dee \Phi=\de \Phi+[A,\Phi]
\quad\in\Omega^1(M,\Omega_\dd^1\aa).\eee
Note that the covariant derivative may be applied to
$\Phi$ thanks to its homogeneous transformation law,
equation (\ref{hom}).

The definition of the Higgs potential in the
infinite dimensional space
\bb V_t(H_t):=(C_t,C_t)\eee
requires a suitable regularisation of the sum of
eigenvalues over the space of spinors $\sss$.
Here, we have to suppose spacetime to be compact and
Euclidean. Then the regularisation is
achieved by the Dixmier trace which allows an explicit
computation of $V_t$.
One of the miracles in CL models is that $V_t$ alone
reproduces the complete bosonic action of a YMH
model. Indeed it consists of three pieces,
the Yang-Mills action, the covariant Klein-Gordon
action and an integrated Higgs potential
\bb V_t(A+H)=\int_M\t (F*F\,z)+ \int_M\t
(\dee\Phi^**\dee\Phi\,z)+ \int_M*V(H).\label{Vt}\ee
As the preliminary Higgs potential $V_0$, the (final)
Higgs potential $V$ is calculated as a function of the
fermion masses,
   \bb V:=V_0-\t[\alpha C^*\alpha C\,z]=
\t[(C-\alpha C)^*(C-\alpha C)\,z].\eee
The linear map
$   \alpha: \Omega_\dd^2\aa\longrightarrow
         \rho(\aa)+\pi(\delta(\ker\pi)^1)$
is determined by the two equations
\bb
\t\lb R^*(C-\alpha C)\,z\rb&=&0\qquad{\rm for\ all}\
R\in\rho(\aa), \label{pj1}\\
 \t\lb K^*\alpha C\,z\rb &=&0\qquad {\rm for\ all}\
K\in\pi(\delta(\ker\pi)^1).\label{pj2}\ee
All remaining traces are over the finite dimensional
Hilbert space $\hh$.

Another miracle happens in the fermionic sector,
where the completely covariant action
 $\psi^*(\dd_t+iH_t)\psi$
reproduces the complete fermionic action of a YMH
model.
We denote by
 \bb\psi=\psi_L+\psi_R\ \in
\hh_t=\sss\,\ot\,\left(\hh_L\op\hh_R\right)\eee
 the multiplets of spinors and by $\psi^*$ the dual of
$\psi$ with respect to the scalar product of the
concerned Hilbert space. Then
\bb\psi^*(\dd_t+iH_t)\psi&=&
\int_M*\psi^*(\ddd+i\gamma(A))\psi
-\int_M*\left(\psi_L^*h\gamma_5\psi_R
+\psi_R^*h^*\gamma_5\psi_L\right)\cr\cr
&&\qq+\int_M*\left(\psi_L^*\mm\gamma_5\psi_R
+\psi_R^*\mm^*\gamma_5\psi_L\right)\cr\cr
&=&\int_M*\psi^*(\ddd+i\gamma(A))\psi
-\int_M*\left(\psi_L^*\varphi\gamma_5\psi_R
+\psi_R^*\varphi^*\gamma_5\psi_L\right)
\label{diract}\ee
containing the ordinary Dirac action
and the Yukawa couplings.
If the minimum $\varphi=v$ is
non-degenerate, we retrieve the input fermionic mass
matrix $\mm$ on the output side by setting the
perturbative variables $h$ to zero in the first equation
in (\ref{diract}). The rhs of the second equation in
(\ref{diract}) is the fermionic action written with the
homogeneous scalar variable $\varphi$. The second
term yields the trilinear invariants (\ref{tri})
with Yukawa couplings fixed such that
$\mm$ is the fermionic mass matrix.
Consequently every CL model is a YMH model with
$\hh_S=\left\{ H\in \Omega_\dd^1\aa,\ H^*=-H
\right\}.$
Note that $\hh_S$ carries a group representation, that
is not necessarily an algebra representation and
we have the following inclusion of group
representations
$
\hh_S\:\subset\:\left(\hh_L^*\ot\hh_R\right)\,\op\,
\left(\hh_R^*\ot\hh_L\right).$

\section{ Necessary conditions}

One may very well do general relativity using only
Euclidean geometry. Still, we agree that
Riemannian geometry is the natural setting of general
relativity. A main argument in favor of this attitude
is that there are infinitely more gravitational theories
in Euclidean geometry than in Riemannian geometry.
The same is true for the standard model. Its natural
setting, to our taste, is non-commutative geometry.
The fact that today's Yang-Mills-Higgs model of
electro-weak and strong interactions falls in the
infinitely smaller class of Connes-Lott models is
remarkable. The purpose of this section is to show in
what extent it is remarkable. We give a list of
constraints on the input of a YMH model. They are
necessary conditions for the existence of a
corresponding CL model.

\subsection{The group}

The compact Lie group $G$ defining a Yang-Mills
model must be chosen such that its Lie algebra $\gg$
admits an invariant scalar product. Therefore $\gg$ is a
direct sum of simple and abelian algebras. After
complexification, the simple Lie algebras are classified
according to E. Cartan, into four infinite series,
$su(n+1),\ n\geq 1,\qq o(2n+1),\ n\geq 2,\qq sp(n),\
n\geq 3,\qq o(2n),\ n\geq 4$ and five exceptional
algebras $G_2,\ F_4,\ E_6,\ E_7,\ E_8$. To define a CL
model, we need a real or complex involution algebra
$\aa$ admitting a finite dimensional, faithful
representation. Their classification is easy. In the
complex case, such an algebra is a direct sum of matrix
algebras $M_n(\cc),\  n\geq 1$. In the real case, we
have direct sums of matrix algebras with real, complex
or quaternionic coefficients, $M_n(\rr),\ M_n(\cc),\
M_n(\hhh),\qq n\geq 1$. The corresponding groups
of unitaries are $O(n,\rr),\ U(n),\ USp(n)$. Note the
two isomorphisms, $USp(2)\cong SU(2)$ and
$USp(4)/\zz_2\cong SO(5,\rr)$.

The groups accessible in a CL model therefore belong
to the second, third, and forth Cartan series.
Furthermore we have $u(n)\cong su(n)\op u(1)$. Up
to the $u(1)$ factor, this is the first series. At the group
level, this factor is disposed of by a condition on the
determinant. In the algebraic setting there is a similar
condition, that reduces the group of unitaries to a
subgroup, here $SU(n)$. This condition is called
unimodularity and is discussed in the next section. To
sum up, all classical Lie groups are accessible in a CL
model but the exceptional ones.

\subsection{The fermion representation}

In a YMH model, the left- and right-handed fermions
come in unitary representations of the chosen group
$G$. Every $G$ has an infinite number of irreducible,
unitary representations. They are classified by their
maximal weight. On the other hand, the above
involution algebras $\aa$ admit only one or two
irreducible representations. The reason is that an
algebra representation has to respect the multiplication
and the linear structure, while a group representation
has to respect only the multiplication. In particular,
the tensor product of two group representations is a
group representation, while the tensor product of two
algebra representations is not an algebra
representation, in general.

 The only irreducible representation of
$M_n(\cc)$ as complex algebra is the fundamental one
on $\hh=\cc^n$. Also $M_n(\rr)$ and $M_n(\hhh)$
have only the fundamental representations on
$\hh=\rr^n$ and $\hh=\cc^n\ot\cc^2$ while
$M_n(\cc)$ considered  as real algebra has two
inequivalent, irreducible representations, the
fundamental one: $\hh=\cc^n$, $\rho_1(a)=a$, $a\in
M_n(\cc)$, and its conjugate: $\hh=\cc^n$,
$\rho_2(a)=\bar a$.

 Therefore, the only possible representations for
fermions in a CL model are
\begin{itemize}
\item
for $G=O(n,\rr)$, $N$ generations of the
fundamental representation on $\hh=\rr^n\ot\rr^N$,
 \item
for $G=U(n)$ (or $SU(n)$ ), $N$ generations of the
fundamental representation on $\hh=\cc^n\ot\cc^N$
and $\bar N$ generations of its conjugate on
$\hh=\cc^n\ot\cc^{\bar N}$,
\item
for $G=USp(n)$, $N$ generations of the
fundamental representation on
$\hh=\cc^n\ot\cc^2\ot\cc^N$.
\end{itemize}

In a YMH model with $G=SU(2)$, the fermions can be
put in any irreducible representations of dimension 1,
2, 3,... while in the corresponding CL model with
$\aa=\hhh$, there is only one irreducible
representation accessible for the fermions, the
fundamental, two dimensional one. Similarly, in a YMH
model with $G=U(1)$ the fermions can have any
(electric) charge from $\zz$ or even from $\rr$ if we
allow `spinor' representations. In the corresponding
CL model with $\aa=\cc$, fermions can only have
charges plus and minus one. In any case, if we want
more fermions in a CL model, we are forced to introduce
families of fermions.

At this point, we realize that all popular grand unified
models are excluded by Connes-Lott.

\subsection{The gauge coupling constants}

In a YMH model, the gauge coupling constants
parameterize the most general gauge invariant scalar
product on the Lie algebra $\gg$ of $G$. In a CL model,
see the rhs of equation (\ref{Vt}), this
scalar product is not general but comes from the trace
over the fermion representation space $\hh$, equation
(\ref{gsp}). The scalar product involves the positive
operator $z$, that commutes with the internal Dirac
operator and with the fermion transformations
$\rho(\aa)$ and that leaves $\hh_L$ and $\hh_R$
invariant. Depending on the details of the mass matrix
and of the left- and  right-handed
 representations $\rho_L$ and $\rho_R$, the
gauge coupling constants may be constraint or not.

\subsection{ The Higgs sector}

As explained in section 2, the scalar representation
$\rho_S$ on $\hh_S$ in a CL model is a representation
of the {\it group} of unitaries only. This
representation is not chosen but it is calculated as a
function of the left- and right-handed fermion
representations and of the mass matrix. The
dependence of the scalar representation on this input
is involved and we can make only one general
statement:
\bb\hh_S\:\subset\:\left(\hh_L^*\ot\hh_R\right)\,\op\,
\left(\hh_R^*\ot\hh_L\right)\eee
which implies that the invariance group of the
fermionic mass terms is equal to the unbroken
subgroup. In a general YMH model the latter is only a
subgroup of the former, e.g. minimal $SU(5)$.
Also, this
inclusion is sufficient to rule out the possibility of
spontaneous parity breaking in left-right symmetric
models \`a la Connes-Lott \cite{is}.

The Higgs potential as well, is on the output side of a CL
model. Its calculation involves the positive
operator $z$ from the input and is by far, the most
complicated calculation in this scheme. We know that
$\varphi=-\mm$ is an absolute minimum of the
Higgs potential. If it is non-degenerate, the gauge and
scalar boson masses are determined by the fermion
masses and the entries of $z$.

Our last necessary condition concerns the Yukawa
couplings. In a CL model, they are determined such that
$\mm$ is the fermionic mass matrix after spontaneous
symmetry breaking. Up to the $z$ dependent scalar
normalization in the bosonic action (\ref{Vt}), the
Yukawa couplings are all one.

\section{The unimodularity condition}

The purpose of the unimodularity condition is to
reduce the group of unitaries $U(n)$ to its subgroup
$SU(n)$. At the group level, this is easily achieved by
the condition $\det g=1$. However the determinant
being a non-linear function is not available at the
algebra level. We are lead to use the trace instead,
together with the formula
\bb \det e^{2\pi iX}\,=\, e^{2\pi i\t X}.\eee
Even in the infinite dimensional case, the connected
component $G^0$ of the unit in the group of unitaries
$G$ is generated by elements $g=e^{2\pi iX}$,
$X=X^*\in\aa$. The desired reduction is achieved by
using the phase, defined by
\bb {\rm phase}_\tau (g):=\frac{1}{2\pi
i}\int_0^1\tau \left(g(t)\frac{\de\ }{\de t}
g(t)^{-1}\right)\,\de t,\eee
 where $\tau$ is a linear
form on $\aa$ satisfying
\bb\tau(1)\in\zz,\qq
\tau(a^*)=\tau(a)^*,\qq \tau(a)=\tau(g^*ag),\qq g\in
G,\  a\in\aa^+:=\{bb^*,\ b\in\aa\},\eee
 and where $g(t)$ is a curve in $G^0$
connecting the unit to $g$. We obtain the finite
dimensional case above by putting
$ \tau(a)=\t \rho(a)$ and $g(t)=e^{2\pi
iXt}$.
The definition of the phase involves two
choices, that are easily controlled in finite dimensions:
the most general linear form $\tau$ can be written as
$\tau(a)=\t \rho (ap),\ a\in\aa,\ p\in{\rm
center}\, \aa,$
and the ambiguity in the choice of the curve $g(t)$ is
controlled by the first fundamental group $\pi^1(G^0)$
which is contained in $\zz$, see table below. Therefore
the unimodularity condition
\bb e^{2\pi i\,{\rm phase}_\tau(g)}=1\eee
is well defined and defines a subgroup
\bb G_p:=\left\{g\in G^0,\
e^{2\pi i\,{\rm phase}_{\t \rho (\cdot
p)}(g)}=1\right\}\eee
 of $G^0$. For $\aa=M_n(\cc)$,
$n\geq 2$, the center is spanned by $1_n$ and
$G_1=SU(n)$. The center of $\aa=M_n(\cc)\op
M_m(\cc)$, $n,m\geq 2$, is spanned by two elements,
$p_n$ and $p_m$,  the projectors on $M_n(\cc)$ and on
$M_m(\cc)$. We have
\bb G_{p_n}&=&SU(n)\times U(m),\cr
       G_{p_m}&=&U(n)\times SU(m),\cr
       G_{p_n+p_m}&=&S(U(n)\times U(m)).\eee

\section{The standard model \`a la
Connes-Lott }

\subsection{Input}

 The standard model in non-commutative geometry is
described by two real algebras, a left-right one for
electro-weak
interactions: $\aa:=\hhh\op\cc$ with group of
unitaries $SU(2)\times U(1)$, and a vector-like
 one for strong
interactions: $\aa':=M_3(\cc)\op\cc$ with group of
unitaries $U(3)\times U(1)$. We denote by $\hhh$ the
algebra of quaternions. Its elements are complex
$2\times 2$ matrices of the form
\bb \pp{x&-\bar y\cr y&\bar x},\qq x,y\in\cc.\eee
Both algebras $\aa$ and $\aa'$  are represented on
the same Hilbert space $\hh=\hh_L\op\hh_R$ of left-
and right-handed fermions,
\bb\hh_L= \left(\cc^2\ot\cc^N\ot\cc^3\right)\ \op\
\left(\cc^2\ot\cc^N\right),\eee
\bb\hh_R=\left((\cc\op\cc)\ot\cc^N\ot\cc^3\right)\
\op\ \left(\cc\ot\cc^N\right).\eee
The first factor denotes weak isospin, the second $N$
generations, $N=3$, and the third denotes colour
triplets and singlets. With respect to the standard basis
\bb \pp{u\cr d}_L,\ \pp{c\cr s}_L,\ \pp{t\cr b}_L,\
\pp{\nu_e\cr e}_L,\ \pp{\nu_\mu\cr\mu}_L,\
\pp{\nu_\tau\cr\tau}_L\eee
of $\hh_L$ and
\bb\matrix{u_R,\cr d_R,}\qq \matrix{c_R,\cr s_R,}\qq
\matrix{t_R,\cr b_R,}\qq  e_R,\qq \mu_R,\qq
\tau_R\eee
 of $\hh_R$, the representations are given
by block diagonal matrices. For $(a,b)\in\hhh\op\cc$
we set
\bb B:=\pp{b&0\cr 0&\bar b}\eee
and define a representation of $\aa$ by
\bb\rho(a,b):=\pp{
a\ot 1_N\ot 1_3&0&0&0\cr
0&a\ot 1_N&0&0\cr
0&0&B\ot 1_N\ot 1_3&0\cr
0&0&0&\bar
b1_N}=\pp{\rho_L(a)&0\cr0&\rho_R(b)}\eee
and for
$(c,d)\in M_3(\cc)\op\cc$ we define a $\aa'$
representation
\bb\rho'(c,d):=\pp{
1_2\ot 1_N\ot c&0&0&0\cr
0&d1_2\ot 1_N&0&0\cr
0&0&1_2\ot 1_N\ot c&0\cr
0&0&0&d1_N}.\eee
The last piece of input is the fermion mass matrix
$\mm$ which constitutes the self adjoint `internal
Dirac operator':
\bb\dd&:=&\pp{
0&0&\pp{M_u\ot1_3&0\cr 0&M_d\ot 1_3}&0\cr
0&0&0&\pp{0\cr M_e}\cr
\pp{M_u^*\ot1_3&0\cr 0&M_d^*\ot 1_3}&0&0&0\cr
0&\pp{0&M_e^*}&0&0}\cr\cr \cr
&=:&\pp{0&\mm\cr\mm^*&0}\eee
with
\bb M_u:=\pp{
m_u&0&0\cr
0&m_c&0\cr
0&0&m_t},\qq M_d:= C_{KM}\pp{
m_d&0&0\cr
0&m_s&0\cr
0&0&m_b},\qq M_e:=\pp{
m_e&0&0\cr
0&m_\mu&0\cr
0&0&m_\tau}\eee
where $C_{KM}$ denotes the
Cabbibo-Kobayashi-Maskawa matrix. All indicated
fermion masses are supposed positive and different.
Note that the
 strong interactions are vector-like: the chirality
operator
 \bb \chi=\pp{
-1_2\ot 1_N\ot 1_3&0&0&0\cr
0&-1_2\ot1_N&0&0\cr
0&0&1_2\ot 1_N\ot 1_3&0\cr
0&0&0&1_N}\eee
and the `Dirac operator' commute with $\aa'$,
$ \left[\dd,\rho'(\aa')\right]=0,$
$ \left[\chi,\rho'(\aa')\right]=0.$

\subsection{Internal space}

We now apply the construction outlined above to the
standard model. Obviously, the standard model is not
the right example to get familiar with the Connes-Lott
scheme. Miraculously enough, the standard
model contains the minimax example, analogue of
the Georgi-Glashow $SO(3)$  model \cite{gg} in the
Yang-Mills-Higgs scheme (a maximum of
pleasure with a minimum of effort). This example
represents the electro-weak algebra $\aa=\hhh\op\cc$
on $two$ generations of leptons. Its only drawback are
neutrinos with electric charge, a drawback, that can be
corrected by adding strong interactions.

Anyway, let us start the computation of the
differential algebra $\Omega_\dd\aa$ for the electro-weak
algebra with
generic element $(a,b)\in\hhh\op \cc$ represented
on the long list of fermions. A general
1-form is a sum of terms
\bb
\pi((a_0,b_0)\delta(a_1,b_1)) =-i \pmatrix
{0&\rho_L(a_0)\left(\mm\rho_R(b_1)-\rho_L(a_1)
\mm\right)\cr \rho_R(b_0)\left(\mm^*\rho_L(a_1)-
\rho_R(b_1) \mm^*\right)&0}\eee
 and as
vector space
\bb   \Omega_\dd^1\aa = \left\{i\pmatrix
{0&\rho_L(h)\mm\cr \mm^*\rho_L(\tilde h^*)&0},\
h,\tilde h\in \hhh\right\}.\eee
The Higgs being an anti-Hermitian 1-form
\bb H=  i\pmatrix
{0&\rho_L(h)\mm\cr \mm^*\rho_L(h^*)&0},\qq
h=\pp{h_1&-\bar h_2\cr h_2&\bar h_1}\in \hhh\eee
is parameterized by one complex doublet
\bb \pp{h_1\cr h_2},\qq h_1,h_2\in\cc.\eee
The junk in degree two turns out to be
\bb J^2=\left\{i\pp{j\ot\Delta&0\cr 0&0},\qq
j\in\hhh\right\}\eee
with
\bb\Delta:=\frac{1}{2}\pp{\left(M_uM_u^*-M_dM_d^*
\right)\ot 1_3&0\cr
0&-M_eM_e^*}.\eee
To project it out, we use the general scalar product
(\ref{gsp}) with the real part of the trace.
Without loss of generality \cite{ks2}, we may
immediately use a $z$, that commutes with
$\rho(\aa)$ {\it and} with $\rho(\aa')$ and of course
with $\dd$ and $\chi$. It  has the form
\bb z=\pp{
x/3\,1_2\ot 1_N\ot 1_3&0&0&0\cr  0&1_2\ot y&0&0\cr
0&0&x/3\,1_2\ot 1_N\ot 1_3&0\cr
0&0&0&y} \label{z}\ee
where $y$ is a positive, diagonal $N\times N$ matrix
and $x$ is a
positive number. The
scalar product defined with this $z$ has a natural
interpretation. Indeed, the general scalar
product is just a sum of the simplest scalar products
in each irreducible part of the fermion
representation, each poised with a separate positive
constant. We have four irreducible parts, the three
lepton families and all quarks together. Due to the
Cabbibo-Kobayashi-Maskawa mixing, the ponderations
of the three quark families are identical. If, in addition,
we suppose that $z$ lie in $\rho$(center$\aa$) then
we have, $y=\lambda 1_N$ with a
positive constant $\lambda$.

 With respect to the general scalar product, we can
write the 2-forms as
\bb \Omega_\dd^2\aa=\left\{\pp{
\tilde c\ot \Sigma &0\cr
0&\mm^*\rho_L(c)\mm},\qq
\tilde c,c\in\hhh\right\}\eee
with
\bb \Sigma:=\frac{1}{2}\pp{\left(M_uM_u^*+M_dM_d^*
\right)\ot 1_3&0\cr
0&M_eM_e^*}.\eee
Since $\pi$ is a
homomorphism of involution algebras, the product  in
$\Omega_\dd\aa$ is given by matrix multiplication
followed by the orthogonal projection $P$ and
the involution  is given by transposition and complex
conjugation. In order to
calculate the  differential $\delta$, we go
back to the universal differential envelope. The result
is
\bb\delta :
\Omega_\dd^1\aa&\longrightarrow&\Omega_\dd^2\aa\cr
\nobreak\cr
i\pmatrix
{0&\rho_L(h)\mm\cr \mm^*\rho_L(\tilde h^*)&0}
 &\longmapsto& \pp{
\tilde c\ot \Sigma&0\cr
0&\mm^*\rho_L(c)\mm}\eee
with
\bb\tilde c=c=h+\tilde h^*.\eee
We are now in position to compute the curvature and
the preliminary Higgs potential:
\bb C:=\delta H+H^2=
\left(1-|\varphi|^2\right)\pp{1_2\ot\Sigma&0 \cr
0&\mm^*\mm}\eee
where we have introduced the {homogeneous} scalar
variable
\bb \Phi:= H-i\dd=:i\pmatrix
{0&\rho_L(\varphi)\mm\cr
\mm^*\rho_L(\varphi^*)&0},\qq
\varphi=\pp{\varphi_1&-\bar \varphi_2\cr
\varphi_2&\bar \varphi_1}\in \hhh,\eee
 \bb |\varphi|^2:=|\varphi_1|^2+|\varphi_2|^2.\eee
The preliminary Higgs potential
 \bb V_0=\t\left[C^2\right]=
 \left(1-|\varphi|^2\right)^2&\times&\left(
\frac{2}{3}\t\left[\left(M_u^*M_u\right)^2\right]x+
\frac{2}{3}\t\left[\left(M_d^*M_d\right)^2\right]x\right.
\cr\cr  &&\qq+
\frac{1}{2}\t\left[M_u^*M_uM_d^*M_d\right]x+
\frac{1}{2}\t\left[M_d^*M_dM_u^*M_u\right]x
\cr \cr&&\left.\qq +
\frac{2}{3}\t\left[\left(M_e^*M_e\right)^2y\right]
\right)\eee
breaks the $SU(2)\times U(1)$ symmetry down to
$U(1)$.

Finally the differential algebra
$\Omega_\dd\aa'$ of the strong algebra is trivial
because strong
interactions are vector-like.

\subsection{Adding spacetime}

 Recall the expression of the
curvature in the electro-weak sector
\bb C=
\left(1-|\varphi|^2\right)\pp{1_2\ot\Sigma&0 \cr
0&\mm^*\mm}.\eee
A straightforward application of equations (\ref{pj1},
\ref{pj2}) --- taking the real part of the traces is
understood --- yields the projection $\alpha C$. It is
again block diagonal with diagonal elements:
\bb \alpha C_{qL}&=&
\frac{1-|\varphi|^2}{2}\frac{
\t\left[M_u^*M_u\right]x+
\t\left[M_d^*M_d\right]x+
\t\left[M_e^*M_e\,y\right]}{Nx+\t y}\,
1_2\ot 1_N\ot 1_3\cr  \cr
\alpha C_{\ell L}&=&
\frac{1-|\varphi|^2}{2}\frac{
\t\left[M_u^*M_u\right]x+
\t\left[M_d^*M_d\right]x+
\t\left[M_e^*M_e\,y\right]}{Nx+\t y}\,
 1_2\ot 1_N\cr  \cr
\alpha C_{qR}&=&
\frac{1-|\varphi|^2}{2}\frac{
\t\left[ M_u^*M_u\right]x+
\t\left[M_d^*M_d\right]x+
\t\left[M_e^*M_e\,y\right]}{Nx+\t y/2}\,
 1_2\ot 1_N\ot 1_3\cr  \cr
\alpha C_{\ell R}&=&
\frac{1-|\varphi|^2}{2}\frac{
\t\left[M_u^*M_u\right]x+
\t\left[M_d^*M_d\right]x+
\t\left[M_e^*M_e\,y\right]}{Nx+\t y/2}\, 1_N.\eee
The Higgs potential is computed next,
\bb V= K\left(1-|\varphi|^2\right)^2,\eee
\bb K&:=&\frac{3}{2}
\t\left[\left(M_u^*M_u\right)^2\right]x+
\frac{3}{2}
\t\left[\left(M_d^*M_d\right)^2\right]x
\cr &&+
\t\left[M_u^*M_uM_d^*M_d\right]x
\cr &&
+\frac{3}{2}
\t\left[M_e^*M_eM_e^*M_e\,y\right]\cr &&
-\frac{1}{2} L^2\left[\frac{1}{Nx+\t y}+
\frac{1}{Nx+\t y/2}\right],\label{k}\ee

\bb L:=\t\left[M_u^*M_u\right]x+
\t\left[M_d^*M_d\right]x+
\t\left[M_e^*M_e\,y\right]. \label{l}\ee
Note that the scalar fields $\varphi_1$ and $\varphi_2$
are not properly normalized, they are dimensionless.
To get their normalization straight we have to compute
the factor in front of the kinetic term in the
Klein-Gordon action:
\bb \t \left(\de\Phi^**\de\Phi\, z\right)=*2L|\partial
\varphi|^2.\eee
Likewise, we need the normalization (cf. appendix of
\cite{is2}) of the electro-weak gauge bosons:
\bb \t \left(F*F\,z\right)=*E
\left(\partial_\mu W_\nu^+\,\partial^\mu
W^{-\nu}-...\right) \eee
with
\bb E:=N x+\t y.\label{e}\ee
We end up with the following masses:
\bb m_W^2&=&\frac{L}{E}, \label{w}\\ \cr
m_H^2&=&\frac{2K}{L}.\label{h}\ee

Finally, we turn to the relations among coupling
constants. As already pointed out,
they are due to the fact that the gauge
invariant scalar product on the internal Lie algebra,
the Lie algebra of the group of unitaries
$ \gg:=\left\{ X\in\aa,\ X^*+X=0\right\}$,
in the Yang-Mills action (\ref{Vt}) is not general
but stems from the trace over the fermion
representation $\rho$ on $\hh$. Since this
representation is faithful the scalar product (\ref{sp})
indeed induces an invariant scalar product on $\gg$.
In the case at hand, our Lie algebra is a direct sum
$\gg\op\gg'$. We define the invariant scalar product
by
\bb \left(X_1+X'_1,X_2+X'_2\right):=
\t\left[\rho(X_1)^*\rho(X_2)\,z\right]+
\t\left[\rho(X'_1)^*\rho(X'_2)\,z'\right]\eee
where $z$ given by equation (\ref{z})
 and $z'$ are two {\it independent} elements in the
intersection of the commutants of $\rho(\aa)$ and
$\rho'(\aa')$,
\bb z'=\pp{
x'/3\,1_2\ot 1_N\ot 1_3&0&0&0\cr  0&1_2\ot y'&0&0\cr
0&0&x'/3\,1_2\ot 1_N\ot 1_3&0\cr
0&0&0&y'} ,\eee
$y'$ is a positive, diagonal $N\times N$ matrix
and $x'$ a
positive number.

The fact that the standard model can be written in the
setting of non-commutative geometry depends
crucially, at this point, on two happy circumstances.
Firstly, the electric charge `generator'
\bb Q=\pp{
\pp{2/3&0\cr 0&-1/3}\ot 1_N\ot 1_3&0&0&0\cr
0&\pp{0&0\cr 0&-1}\ot 1_N&0&0\cr
0&0&\pp{2/3&0\cr 0&-1/3}\ot 1_N\ot 1_3&0\cr
0&0&0&-1_N}\eee
is an element of $i\rho(\gg)\op i\rho'(\gg')$. Indeed it
 is a linear
combination of weak isospin $I_3$ and elements of the
three $u(1)$ factors:
\bb Q=\rho\left(\pp{1/2&0\cr 0&-1/2},0\right)+
\frac{1}{2i}\rho(0,i)+\frac{1}{6i}\rho'(i1_3,0)
-\frac{1}{2i}\rho'(0,i).\eee
We have put `generator' in quotation marks because
 $iQ$ is a Lie algebra element, not $Q$.
The weak angle $\theta_w$ measures the proportion
of weak isospin in the electric charge:
\bb \frac{Q}{|Q|}= \sin\theta_w\,\frac{I_3}{|I_3|}
+\cos\theta_w\,\frac{Y}{|Y|}.\label{sin}\ee
The hypercharge $Y$ is a linear combination of the
three $u(1)$ factors,
\bb Y:=\frac{1}{2i}\rho(0,i)+\frac{1}{6i}\rho'(i1_3,0)
-\frac{1}{2i}\rho'(0,i).\eee
Here comes the second happy circumstance,
this particular combination $Y$ is singled out by two
unimodularity conditions. They reduce the
group of unitaries $SU(2)\times U(1)\times U(3)\times
U(1)$ to $SU(2)\times U(1)\times SU(3)$ with the
surviving $U(1)$ generated by the hypercharge.
Indeed, the center of $\aa\op\aa'$ is four dimensional
with basis $p_1,...,p_4$. $p_1:=\rho(1_2,0)$
projects on $\hhh$, $p_2:=\rho(0,1)$ on $\cc$,
$p_3:=\rho'(1_3,0)$ on $M_3(\cc)$, and
$p_4=\rho'(0,1)$ on $\cc'$, and the group of the
standard model is $G_{p_1}\cap G_{p_2}$.

Let us come back to the calculation of the weak angle.
 Equation (\ref{sin}) is a matrix of equations. Let
us take the difference of the two diagonal elements
corresponding to the left handed neutrino and electron:
\bb \frac{1}{|Q|}=\sin\theta_w\,\frac{1}{|I_3|},\eee
\bb \sin^2\theta_w=\frac{(I_3,I_3)}{(Q,Q)}.\eee
The numerator is readily computed,
\bb (I_3,I_3)=
\t\left[\rho\left(\pp{1/2&0\cr 0&-1/2},0\right)^2
\,z\right]=\frac{1}{2}(Nx+\t y).\eee
We compute the denominator with Pythagoras' kind
help,
\bb (Q,Q)&=&\t\left[\rho\left(\pp{1/2&0\cr 0&-1/2}
,0\right)^2\,z\right]+
\frac{1}{4}\t\left[\rho(0,1)^2\,z\right]\cr
&&+\frac{1}{36}\t\left[\rho'(1_3,0)^2\,z'\right]
+\frac{1}{4}\t\left[\rho'(0,1)^2\,z'\right]\cr \cr
&=&(Nx+\frac{3}{4}\t y)
+\frac{1}{3}x'+\frac{3}{4}\t y'.\eee
Finally the mixing angle is given by
\bb \sin^2\theta_w=\frac{Nx+\t y}
{2Nx+\frac{3}{2}\t y
+\frac{2}{3}x'+\frac{3}{2}\t y'}
.\label{s}\ee
In a similar fashion, the ratio
between strong and weak coupling is computed,
\bb \left(\frac{g_3}{g_2}\right)^2=
\frac{(I_3,I_3)}{(C,C)}=\frac{1}{4}\frac{Nx+\t y}
{x'} \label{g}\ee
where
\bb C:=\rho'\left(\pp{1/2&0&0\cr 0&-1/2&0\cr
0&0&0},0\right).\eee
Here $C$ stands for colour not for curvature.

In this calculation $z$ and $z'$ are different in
general, implying that the electro-weak sector
$\rho(\aa)$ is orthogonal to the strong sector
$\rho'(\aa')$. In the special case where $z=z'$ a
different choice is possible:
\bb (a,a'):=\t \left[\rho(a)^*\rho'(a')\,z\right],\qq
a\in\aa,\ a'\in\aa'.\eee
Then the two $U(1)$ factors $\rho(0,1)$ and
$\rho'(0,1)$ are not orthogonal anymore and the value
of $\sin^2\theta_w$ comes out smaller \cite{ks}. This
choice is closer to grand unified models and yields
$\sin^2\theta_w=3/8=0.375$ for $z=z'=1$ to be compared
to $\sin^2\theta_w=12/29=0.414$ from equation
(\ref{s}).

\section{Fuzzy relations}

Non-commutative geometry
produces relations among gauge couplings and
boson and fermion
masses. The aim of this section is a detailed study of
these relations for the standard model. Here, we will
encounter a new phenomenon, that we call fuzzy
relations. To get a feeling for this phenomenon, it will
be helpful to consider first simpler models. We start by
switching off strong interactions. Indeed, since they
are vector-like, they do not yet play an important role
in the non-commutative setting.

So let us consider the real algebra $\hhh\op\cc$
with $N$ generations of leptons. In this case
equations (\ref{e}, \ref{k}, \ref{l}) reduce to
\bb E&=&\sum_{j=1}^N y_j,\cr
 K&=&\frac{3}{2}\,\sum m^4_jy_j\,-\,
\frac{1}{2}\,L^2\left[\frac{1}{\sum
y_j}+\frac{2}{\sum y_j}\right]
,\cr\cr  L&=&\sum m_j^2y_j,\eee
where the $y_j$ denote the eigenvalues of the positive
diagonal matrix $y$. Recall that they
are arbitrary positive parameters.
 Let us also recall
the expressions (\ref{w}, \ref{h}) of the $W$ and
Higgs masses
$m_W^2=L/E,\qq
m_H^2=2K/L$.
Since only squares of masses appear, we will alleviate
notations by putting
\bb m_W^2=:W,\qq m_H^2=:H,\qq m_\tau^2=:\tau,\qq ...
\eee
Now, one generation,
$N=1$, has a degenerate vacuum, i.e. a vanishing Higgs
potential, $K=0$. For more than one generations, this
degeneracy is lifted.

If we take two generations, say $\mu<\tau$, we
can eliminate the two positive unknown  $y_j$
 from equations (\ref{w}, \ref{h}) and we obtain
the {\it exact} mass relation
\bb H=3\left(\tau-W\right)\left(1-
\frac{\mu}{W}\right)\label{lex}\ee
 with $ \mu\,<\,W\,<\,\tau$.
This curve in the
$m_\tau\,m_H$ plane
is again a degenerate
situation in the sense that with $N=3$ generations
(or more),
$e<\mu<\tau$, this curve will become
a band of width
\bb\sqrt{\frac{m_\mu^2-m_e^2}{m_W^2}}\,\,
\sqrt{3\left(m_\tau^2-m_W^2\right)}.\eee
 This is what we call a {\it fuzzy} mass relation.

Here are the details for $N=3$. Equations
(\ref{w}, \ref{h}) are homogeneous in our three
positive unknowns $y_j$. Therefore we introduce
\bb z_1:=y_1/y_3,\qq z_2:=y_2/y_3,\eee
and solve equation (\ref{w}) with respect to
$z_2$,
\bb z_2=\frac{\tau-W-z_1(W-e)}{W-\mu},\eee
 Eliminating
$z_2$ from equation (\ref{h}) we get
\bb H/3+W=\frac{-z_1(\mu-e)[\mu+e\,-\mu e/W]
+(\tau-\mu)[\tau+\mu\,-\tau\mu/W]}
{-z_1(\mu-e)+(\tau-\mu)}\label{hl}.\ee
{}From equation
(\ref{w}), we know that the $W$ mass lies between
the masses of the lightest and of the heaviest lepton,
$e<W<\tau$. Therefore, we have to distinguish
two cases, $\mu< W$ and $\mu>W$. In the first case, as
$z_2$ is positive, $z_1$ varies in a finite interval
\bb 0<z_1<\frac{\tau-W}{W-e}.\label{mm}\ee
On the other hand, one checks easily that the rhs of
equation (\ref{hl}) is an increasing function of $z_1$,
and the inequalities (\ref{mm})
imply the inequalities
\bb m_H(m_\tau;m_\mu)\,<\,m_H\,<\,
m_H(m_\tau;m_e),\label{ban}\ee
where we introduced the parameterized family of
curves in the
$ m_\tau\,m_H$ plane
\bb m_H(m_\tau;m):=
\sqrt{3\left(m_\tau^2-m_W^2\right)
\left(1-m^2/m_W^2\right)}.\eee
The parameter $m$ varies in the open interval
$(0,m_W)$.
All values of $m_H$ in the open interval described by
$m\in (m_e,m_\mu)$ do occur. In the degenerate case
$m_e=m_\mu$, the band (\ref{ban}) collapses to the
curve $m_H(m_\tau;m_\mu)$ which is the graph
corresponding to two generations, equation
(\ref{lex}).

 In the second case, $W<\mu$, $z_1$ varies in an
infinite interval,
\bb \frac{\tau-W}{W-e}<z_1<\infty,\eee
and the Higgs mass is now a decreasing function of
$z_1$. Again, we get two inequalities,
\bb 3(W-e)\left(\frac{\mu}{W}-1\right)\,<\,H
\,<\,3(W-e)\left(\frac{\tau}{W}-1\right),\eee
that reduce to the degenerate case, equation
(\ref{lex}), for $\mu=\tau$. Note that these
inequalities remain valid for $\mu=W$.

Let us now consider the relation among the gauge
couplings $g_2$ and $g_1$ in the $\hhh\op\cc$
model. For leptons only and any number of
generations, we have the {\it exact} relation \cite{is2}
\bb g_2= \sqrt{2}\,g_1,\qq\sin^2\theta=1/3.\eee
If we require the relations among gauge couplings to
be fuzzy as well, we must add at least one generation
of quarks. Then we get
\bb 1/5\,<\,\sin^2\theta\,<\,1/3.\label{sinl}\ee
Note that, if we admit right handed neutrinos,
$\sin^2\theta_w=1/5$ and it can not be made fuzzy by
the addition of quarks.

The analysis of the fuzzy mass relations in the
presence of quarks is more complicated than
the purely leptonic analysis above. To simplify, let us
take two generations of leptons, $\mu$ and $\tau$, and
one generation of quarks, $t$ and $b$.
Then equations (\ref{e}, \ref{k}, \ref{l}) read\
\bb E&=&x+y_2+y_3,\cr
 K&=&\frac{3}{2}\,(t^2+b^2)x+\,tbx\,+\,
\frac{3}{2}\,(\mu^2y_2+\tau^2y_3)\cr
&&\,-\,\frac{1}{2}\,L^2\left[\frac{1}{x+y_2+y_3}+
\frac{1}{x+(y_2+y_3)/2}\right],\cr \cr
 L&=&(t+b)x\,+\,\mu y_2+\tau y_3.\ee
Even after this simplification,
many different cases have to be distinguished whereas
in the last example we only had two cases. Let us only
treat one case here: if we assume $\mu<\tau<(1-
1/\sqrt{3})(t+b)$ then the Higgs mass will also be a
monotonic function, just as in the three
lepton example.
As before, equation (\ref{w}) yields
lower and upper bounds for the $W$ mass,
$ e<W<t+b.$
Adapting the analysis of three lepton generations to
the present case, we find again that the Higgs mass
varies in a finite, open interval, $H_{\rm
min}<H<H_{\rm max}$ with
\bb H_{\rm min}&=&
3\left(t+b+\tau-(t+b)\frac{\tau}{W}\right)
\,-\,W\,\frac{3(t+b)+W-4\tau}{t+b+W-2\tau}
\,-\, 4t\,\frac{b}{W}\,\frac{W-\tau}{t+b-\tau},
\label{hmin}\\ \cr
H_{\rm max}&=&
3\left(t+b+\mu-(t+b)\frac{\mu}{W}\right)
\,-\,W\,\frac{3(t+b)+W-4\mu}{t+b+W-2\mu}
\,-\, 4t\,\frac{b}{W}\,\frac{W-\mu}{t+b-\mu}
\label{hmax}.\ee
 Away from the lower bound $m_W^2-m_b^2$ of
$m_t^2$, the width of the allowed band in the
$m_t\,m_H$ plane is again governed by the (light)
leptons, in the sense that the band collapses if
$m_\tau=m_\mu$.

Let us put back colour and consider the standard
model with $N=3$ generations of leptons and quarks.
Note that now in equations (\ref{e}, \ref{k}, \ref{l}),
 all three
quark generations are poised with the same positive
parameter $x$, while the three lepton generations are
poised independently with the three positive
parameters $y_1,\ y_2$ and $y_3$:
\bb
E&=&3x+y_1+y_2+y_3,\label{cc}\\
K&=&\frac{3}{2}\,(u^2+d^2+c^2+s^2+t^2+b^2)x+
\,(ud+cs+tb)x\,\cr &&+\,
\frac{3}{2}\,(e^2y_1+\mu^2y_2+\tau^2y_3)\cr
&&\,-\,\frac{1}{2}\,L^2\left[\frac{1}{3x+y_1+y_2+y_3}+
\frac{1}{3x+(y_1+y_2+y_3)/2}\right],\\ \cr
 L&=&(u+d+c+s+t+b)x\,+\,ey_1+\mu y_2+\tau
y_3.\label{kk}\ee
Recall that this
difference is due to the quark mixing given by a
non-degenerate Cabbibo-Kobayashi-Maskawa matrix.
Here non-degenerate means that there are no common
mass and weak interactions eigenstates in the quark
sector. This reduction of parameters modifies the
bounds on the $W$ mass,
  \bb e\,<\,W\,<\,(u+d+c+s+t+b)/3\eee
otherwise the
Cabbibo-Kobayashi-Maskawa matrix drops out
of equations (\ref{cc}-\ref{kk}). Note that colour does
not affect the $W$ and Higgs masses because of
the vector character of strong interactions and
because of the homogeneous appearance of the
parameters $x,\ y_1,\ y_2,\ y_3$ in equations
(\ref{w}, \ref{h}). In particular, we get lower and
upper bounds on the Higgs mass similar to equations
(\ref{hmin}) and (\ref{hmax}) if we restrict ourselves
here to the case  $\tau\,<\,t+b$. Again, putting
the lepton masses to zero makes these bounds collapse,
the fuzzy mass relation becomes exact,
 \bb
H=\frac{3(u^2+d^2+c^2+s^2+t^2+b^2)+2(ud+cs+tb)}
{u+d+c+s+t+b}-3W\,\frac{u+d+c+s+t+b+W}
{u+d+c+s+t+b+3W}.\eee

 The complete analysis will be published elsewhere
\cite{iks}.

Finally let us discuss the relations among gauge
couplings in the standard model. The addition of
colour changes the picture quite drastically
 because of
the additional element $z'$ in the commutants and
because of the strong gauge coupling $g_3$. Recall the
gauge coupling ratios
\bb \sin^2\theta_w=\frac{3x+\t y}
{6x+\frac{3}{2}\t y+\frac{2}{3}x'+\frac{3}{2}\t
y'},\eee and
\bb \left(\frac{g_3}{g_2}\right)^2=\,
\frac{3x+\t y}{4x'}.\eee
Consequently, the strong gauge coupling is
arbitrary. This is natural. However, via the
unimodularity condition, it back reacts on the weak
mixing angle and
\bb\sin^2\theta_w&<&\frac{2}{3}\,
\left(1+\frac{W}{u+d+c+s+t+b}+\frac{1}{9}\,
\left(\frac{g_2}{g_3}\right)^2\right)^{-1}
\qq {\rm if}\ e=\mu=\tau=0.\eee
Numerically, this back reaction is negligible,
$(g_2/g_3)^2=0.015$. However, for non-zero lepton
masses, even for light leptons, the optimal bound
 of $\sin^2\theta_w$ reduces to 2/3 annihilating the
mentioned back reaction.

For the natural subclass of scalar products defined
with $z$ and $z'$ in
$\rho($center $\aa)\cap\rho'($center $\aa')$ we have
\bb y=\frac{x}{3}\,\,1_3,\qq
y'=\frac{x'}{3}\,\,1_3.\eee
Consequently, the fuzziness of the mass relations is lost,
\bb W&=&(u+d+c+s+t+b)/4\,+\,(e+\mu+\tau)/12,\cr\cr
H&=&\frac{3(u^2+d^2+c^2+s^2+t^2+b^2)
\,+\,2(ud+cs+tb)\,+\,(\tau^2+\mu^2+e^2)}
{4W}\,-\,\frac{15}{7}W.\eee
The ratios of gauge couplings reduce to
\bb
\sin^2\theta_w=\frac{24}{45+13(g_2/g_3)^2}\,<\,
\frac{8}{15},\eee
\bb\left(\frac{g_3}{g_2}\right)^2\,=\,\frac{x}{x'}.\eee

 In
conclusion, as a Yang-Mills-Higgs model, the standard
model can be accommodated in the very narrow frame
of non-commutative geometry under two conditions.
The first condition concerns the representation
content, fermions must sit exclusively in fundamental
or singlet representations and the Higgs scalar sits in
{\it one} weak isospin doublet. The second condition
concerns gauge couplings and masses and we find a
rich structure. The Higgs mass is determined by all
fermion masses with a conceptual  uncertainty of one
part per thousand, $m_H =280\pm 33\ GeV$ for
$m_t=176\pm 18\ GeV$. Naturally, we interpret this
prediction to hold for pole masses, because the pole
masses are gauge invariant. Nevertheless, should the
reader be inclined to interpret the relations among
masses and gauge couplings at the scale dependent
level, then he may do so. Indeed, their inherent
'fuzziness' renders them stable under local
renormalization flow and this should be enough as the
theory does not contain any super heavy scale
\cite{ren}. In any
case, this rich structure deserves further theoretical
and experimental exploration.

\vskip1truecm

It is as pleasure to acknowledge Alain Connes' helpful
advice.


\begin{thebibliography}{77}
\bibitem{c}
 A. Connes, {\it Non-Commutative Geometry}, Academic
Press (1994)
\bibitem{cl}
 A. Connes \& J. Lott, {\it The metric
aspect of non-commutative geometry}, in the
proceedings of the 1991 Carg\`ese Summer Conference,
eds.: J. Fr\"ohlich et al., Plenum Press (1992)
\bibitem{ks2}
D. Kastler \& T. Sch\"ucker, {\it The standard model \`a
la Connes-Lott}, CPT-94/P.3091, hep-th/9412185 (1994)
\hfil\break
D. Kastler \& T. Sch\"ucker, {\it A detailed account of Alain
Connes' version of the standard model in
non-commutative geometry, IV}, CPT-94/P.3092 (1994),
hep-th/9501077
\bibitem{is2} B. Iochum \& T. Sch\"ucker, {\it
Yang-Mills-Higgs versus Connes-Lott},
hep-th/9501142, Comm. Math. Phys., to appear
\bibitem{k}
D. Kastler, {\it A detailed account of Alain
Connes' version of the standard model in
non-commutative geometry, I and II}, Rev. Math. Phys.
5 (1993) 477 \hfil\break
D. Kastler, {\it A detailed account of Alain
Connes' version of the standard model in
non-commutative geometry, III}, CPT-92/P.2824
(1992)  \hfil\break
D. Kastler \& M. Mebkhout, {\it Lectures on
Non-Commutative Differential Geometry}, World
Scientific, to be published
\bibitem{vg}
J. C. V\'arilly \&  J. M. Gracia-Bond\'\i a,
{\it Connes' noncommutative differential geometry and
the standard model}, J. Geom. Phys. 12 (1993) 223
\bibitem{orsay}
 M. Dubois-Violette, {\it D\'erivations et calcul
diff\'erentiel non commutatif},
 C. R. Acad. Sc. Paris 307 I (1988) 403
\hfil\break
 M. Dubois-Violette, R. Kerner \& J.
Madore,   {\it Classical bosons in a non-commutative
geometry},
 Class. Quant. Grav. 6 (1989) 1709,   {\it Noncommutative
differential geometry of matrix algebras}, J. Math.
Phys 31 (1990) 316,   {\it Noncommutative differential
geometry and new models of gauge theory}, J. Math.
Phys. 31 (1990) 323
\hfil\break
J. Madore, {\it An Introduction to Noncommutative
Differential Geometry and its Physical Applications},
Cambridge University Press (1995)
\bibitem{mm}
 R. Coquereaux, G. Esposito-Far\`ese \&
G. Vaillant,  {\it Higgs fields as Yang-Mills fields and
discrete symmetries}, Nucl. Phys. B353 (1991) 689
\hfil\break
 R. Coquereaux, G. Esposito-Far\`ese \&
F. Scheck,  {\it Noncommutative geometry and graded
algebras in electro\-weak interactions},
 Int. J. Mod. Phys. A7 (1992) 6555
\hfil\break
R. Coquereaux, R. H\"au\ss ling, N.A. Papadopoulos  \&
F. Scheck,  {\it Generalized gauge transformations and
hidden symmetries in the standard model},
 Int. J. Mod. Phys. A7 (1992) 2809
\hfil\break
N.A. Papadopoulos, J. Plass \& F. Scheck,  {\it Models of
electroweak interactions in non-commutative
geometry: a comparison}, Phys. Lett. 323B (1994) 380
\bibitem{zur}
 A. Chamseddine,
G. Felder \& J. Fr\"ohlich,   {\it Grand unification in
non-commutative geometry}, Nucl. Phys. B395 (1993)
672 \hfil\break
 A. Chamseddine \& J. Fr\"ohlich,  {\it
$SO(10)$ Unification in non-commutative geometry},
Phys. Rev. D50 (1994) 2893
\bibitem{kppw}
W. Kalau, N. A. Papadopoulos, J. Plass \& J.-M.
Warzecha, {\it Differential algebras in
non-commutative geometry}, J. Geom. Phys., in press
\bibitem{sz}
T. Sch\"ucker \& J.-M. Zylinski, {\it Connes' model
building kit}, J. Geom. Phys. 16 (1994) 1
\bibitem{is}
B. Iochum \& T. Sch\"ucker, {\it A left-right symmetric
model \`a la Connes-Lott}, Lett. Math. Phys. 32 (1994)
153
\bibitem{gg}
H. Georgi \& S. L. Glashow, {\it Unified weak
and electromagnetic interactions without
neutral currents}, Phys. Rev. Lett. 28 (1972) 1494
\bibitem{ks}
D. Kastler \& T. Sch\"ucker, {\it Remarks on Alain
Connes' approach to the
standard model in non-commutative geometry},
 Theor. Math. Phys. 92
(1992) 522, English version, 92 (1993) 1075
\bibitem{iks}
B. Iochum, D. Kastler \& T. Sch\"ucker
{\it  Fuzzy mass relations in the standard model}, in
preparation \bibitem{ren}
J. Kubo, K. Sibold \& W. Zimmermann,
{\it Higgs and top mass from reduction of couplings},
Nucl. Phys. B259 (1985) 331\hfil\break
E. Alvarez, J. M. Gracia-Bond\'\i a \& C. P. Mart\'\i n,
{\it A renormalisation group analysis of the NCG
constraints $m_{top}=2\,m_{W}$,
$m_{Higgs}=3.14\,m_{W}$}, Phys. Lett. 323 (1994) 259

 \end{thebibliography}
 \end{document}